# Latest Results on Cavity Gradient and Input RF Stability at FLASH/TTF Facility

Shilun Pei[1], Chris E. Adolphsen[1], John Carwardine[2] and Nicholas John Walker[3] [*]

1 – SLAC National Accelerator Laboratory
2575 Sand Hill Road, Menlo Park, CA 94025 – United States

2 – Argonne National Laboratory (ANL)
9700 S. Cass Avenue, Argonne, IL 60439 – United States

3 – DESY-Hamburg site, Deutsches Elektronen-Synchrotoron in der Helmholtz-Gemeinschaft, Notkestrasse 85, 22607 Hamburg – Germany

The FLASH L-band (1.3 GHz) superconducting accelerator facility at DESY has a Low Level RF (LLRF) system that is similar to that envisioned for ILC. This system has extensive monitoring capability and was used to gather performance data relevant to ILC. Recently, waveform data were recorded with both beam on and off for three, 8-cavity cryomodules to evaluate the input RF and cavity gradient stability and study the rf overhead required to achieve constant gradient during the 800µs pulses. In this paper, we present the recent experimental results and discuss the pulse-to-pulse input rf and cavity gradient stability for both the beam on and off cases. In addition, a model of the gradient variation observed in the beam off case will be described.

## 1  Introduction

The FLASH facility at DESY is the world's only FEL for VUV and soft X-ray production. Its layout is shown in Figure 1. Presently, it includes six accelerator modules each containing eight, L-band (1.3 GHz), 1-m long, 9-cell, superconducting cavities. The three modules, ACC4-ACC6, are the focus of this study as they are very similar to an ILC rf unit. These 24 cavities are powered by a single klystron and the LLRF system monitors the input and reflected rf at each cavity as well as the cavity fields using probe couplers. The probe signals for the 24 cavities are summed vectorally and used by the LLRF feedback system to keep the net gradient from the 24 cavities constant during the 800µs beam period that follow a 500µs cavity fill period.

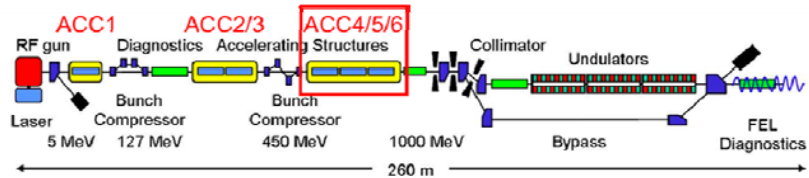

Figure 1: Layout of the FLASH facility.

The LLRF performance study is part of the '9mA' program in support of FLASH, XFEL and ILC [2]. There have been four periods of 9mA studies to date, in May 2008, September 2008, January 2009 and September 2009. The study of cavity gradient and input rf stability was started in September 2008 when three sets of the data were taken in which: 1) Feedback and Adaptive Feed Forward were off; 2) Feedback was on and Adaptive Feed Forward was off; and 3) Feedback and Adaptive Feed Forward were on - the study results can be found in Reference [3]. Based on these results, dedicated open-loop, beam-off experiments were done in January 2009; the experimental results further confirmed the key findings in September 2008. In September 2009, stable Feedback-on operation with high beam loading was demonstrated [4], but routine operation of long bunch trains will require further improvements.

---

[*] Work supported by the DOE under contract DE-AC02-76SF00515.

Here we discuss the measurements and analysis results from data taken during the last two runs. Analytical estimates of the gradient jitter are also presented.

## 2 Beam, Feedback and Adaptive Feed Forward off studies

For the experiments in January 2009, the accelerator gradient for cavities in ACC4-ACC6 was varied for three cavity frequency settings (nominal, but not necessarily optimal, and +100 Hz and -100 Hz relative to nominal). Also, data were taken with piezo actuator compensation of the Lorentz force detuning in the cavities of module ACC6. The beam, Feedback and Adaptive Feed Forward were off in all cases except the Feedback was turned on when the piezos were used.

### 2.1 Input rf signals

Figure 2 shows the input rf jitter for different input rf amplitudes in two periods: during the fill (1st flat top) and during the nominal beam period (2nd flat top). The error bars indicate the range of jitter in each period. The jitter is roughly inversely proportional to the input rf amplitude, which means the jitter likely originates from noise in the rf drive and the absolute rf jitter is roughly constant. The 2nd flat top jitter is twice that of the 1st because the amplitude is halved during the 2nd period for beam-off operation. The cavity detuning has no effect on the input rf jitter as expected for operation with Feedback off. No significant reflected-to-forward rf cross-talk was observed as in earlier runs due to the improvement of the power coupler directionality.

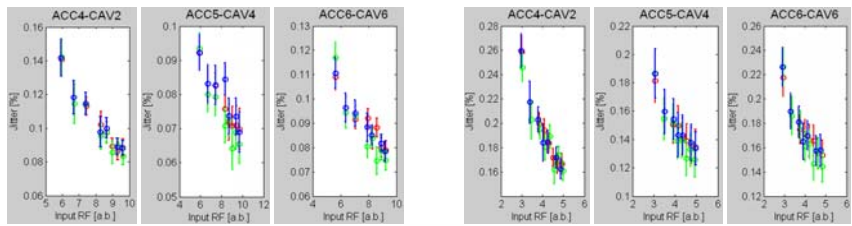

Figure 2: 1st (left three) and 2nd (right three) input rf flat top jitter as a function of rf amplitude. Red: nominal frequency; blue: nominal +100Hz; green: nominal-100Hz.

### 2.2 Cavity probe signals

Figure 3 shows the measured probe signal jitter for different cavity gradients (the probe signal is proportional to the cavity gradient). It can be seen that for each cavity and detuning, there is an optimal operating gradient with minimum probe jitter. This is also seen in the analytical model estimation [5, 6] results shown in Figure 4. Here the measured input rf signals were used and a 3 Hz Gaussian pulse-to-pulse detuning jitter was included to simulate the effect of microphonics. With increased detuning, the optimal cavity gradient increases as larger Lorentz forces, which lower the cavity frequency, are required to bring the cavity back in tune where the jitter is minimal. Figure 5 shows the simulated probe signal jitter as a function of detuning frequency for the specified cavity gradient.

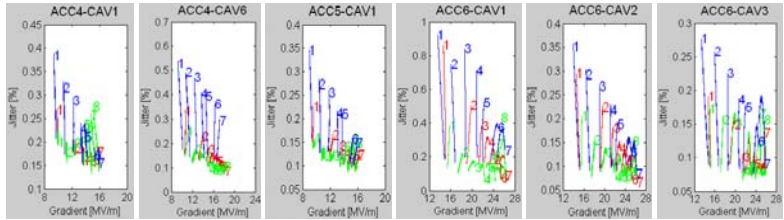

Figure 3: Jitter along the probe signal flat top as a function of cavity gradient. The numbers refer to the data set and the colors refer to the detuning as in Figure 1.

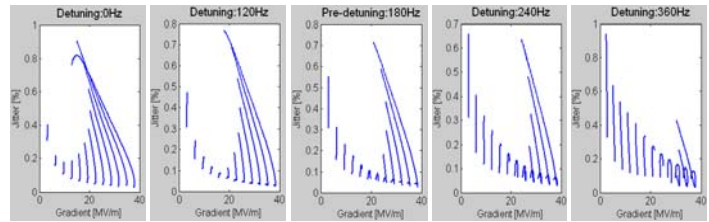

Figure 4: Analytical estimates of the probe signal flat top jitter versus cavity gradient for the specified detuning.

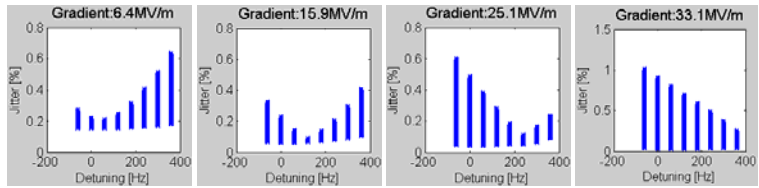

Figure 5: Analytical estimates of the probe signal flat top jitter versus detuning for the specified cavity gradient.

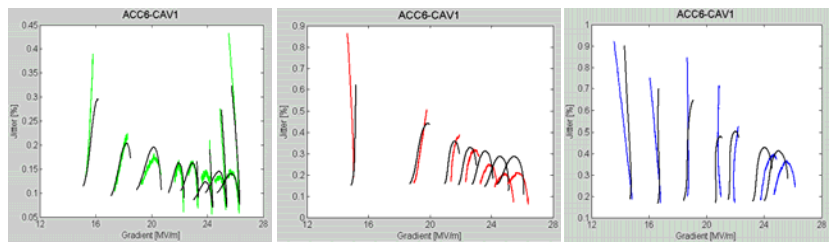

Figure 6: Comparison between data and simulated probe jitter for different detuning. Green: nominal-100Hz; red: nominal; blue: nominal+100Hz; black: simulation.

Figure 6 shows a comparison between data and simulation of probe signal flat top jitter in ACC6-CAV1. Only two variables, the cavity detuning (110Hz, 190Hz and 240Hz from left to right) and the jitter on that detuning (8Hz, 7Hz and 8Hz rms from left to right), were used to match the data. The good agreement indicates that the jitter is dominated by the initial (pre-pulse) detuning, likely from microphonics.

Figure 7 shows the probe signal flat top jitter for all 24 cavities in ACC4-ACC6 at all detuning and gradient settings. The jitter is well below 1% for all gradients and generally decreases with gradient (reflecting the choices of cavity frequencies). This differs from the results described in Reference [3] in which up to 4% jitter was observed for specific cavities at higher gradient. The reflected-to-forward rf ratio is ~50% at the end of the 1st flat top in the earlier data, while it is about ~20% for the current data, which indicates the cavity frequencies were better optimized (this ratio is ideally zero but without Lorentz force detuning compensation, it will be finite).

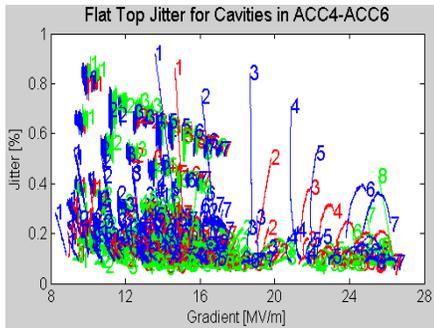 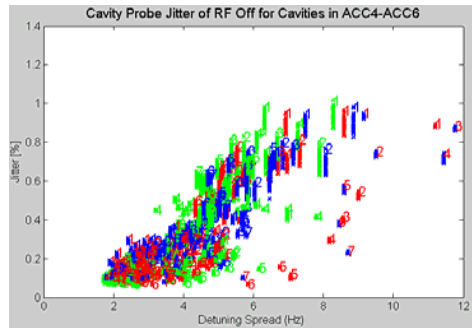

Figure 7: Probe signal jitter for all cavities/detunings.

Figure 8: Probe signal jitter at the end of the rf pulse versus detuning jitter for all cavities/detunings.

Figure 8 shows that there is a strong correlation between the measured probe signal jitter and detuning jitter at the end of rf pulse, as was the case with earlier data [3] (the frequency of the rf discharged from the cavity after the pulse was used to determine the cavity detuning). This correlation is consistent with the analytical model discussed above.

Figure 9 shows the probe signal flat top jitter with the piezos on and off for cavities in ACC6 with nominal detuning. As expected, the reflected-to-forward rf ratio was greatly reduced with piezo compensation (Feedback was on), from about 20% to less than 5% [6]. However, it appears that the piezo actuators introduced additional gradient jitter in some cases.

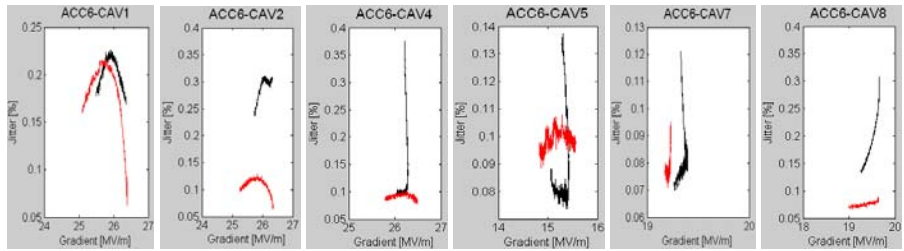

Figure 9: Probe signal flat top jitter with piezos on and off for nominal detuning. Red: piezos off; black: piezos on.

## 3   Heavy beam loading experiments

The heavy beam loading experiments were carried out on September 17-20, 2009. Adaptive Feed Forward and orbit feedback were not tried, only Feedback was used, and both piezos on and off data for ACC6 were collected. 3mA (1MHz/3nC) beam currents ran stably for trains of 800 bunches (800µs); 9mA (3MHz/3nC) currents ran stably for trains of 900-1500 bunches (300-500µs) but with some mid-train trips from beam losses; 9mA (3MHz/3nC) currents ran with close to 2400 bunches (800µs) with many mid-train trips.

### 3.1   Input rf signals

Figure 10 shows the input rf flat top jitter for beam-on data. With the absence of the beam during the 1st flat top period, the jitter is at the same level as the beam off, open loop case. However, the 2nd flat top jitter is much larger as a result of the Feedback system compensating for beam current jitter.

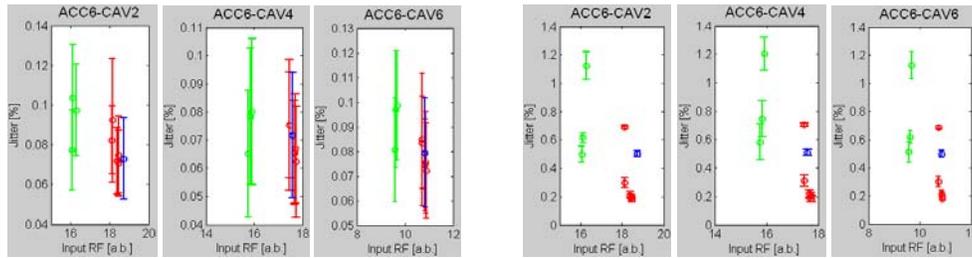

Figure 10: 1st (left three) and 2nd (right three) input rf flat top jitter as a function of input rf amplitude. Red: 1MHz/3nC beam with piezos off; Blue: 3MHz/3nC beam with piezos off; Green: 3MHz/3nC beam with piezos on.

Figure 11 shows the 2nd input rf flat top waveforms for the four cavities with the highest input rf power in the whole rf unit (ACC4-ACC6). Comparing the right and left plots shows the piezo actuators reduced the required rf overhead (i.e., made the waveforms flatter - having a full piezo compliment should make them even flatter). This overhead would be even smaller if all 24 of the cavities in the whole rf unit had piezo compensation instead of 8 in this case.

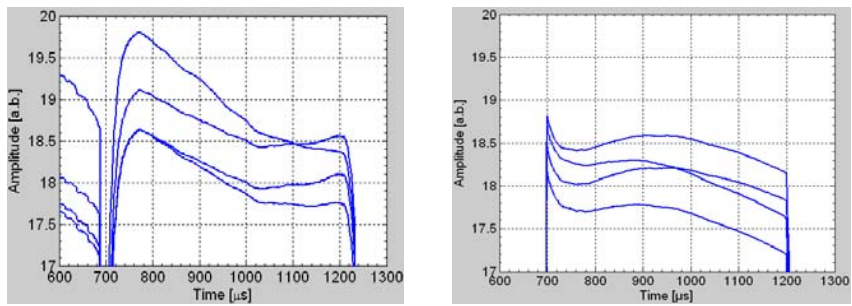

Figure 11: 2nd input rf flat top waveforms. Left: 3MHz/3nC beam with 1600 bunches and piezos off; Right: 3MHz/3nC beam with 1500 bunches and piezos on.

Figure 12 shows that the 2nd input rf flat top jitter is linearly correlated with the bunch charge jitter. The slopes agree well with those expected based on the beam loading fraction (solid lines).

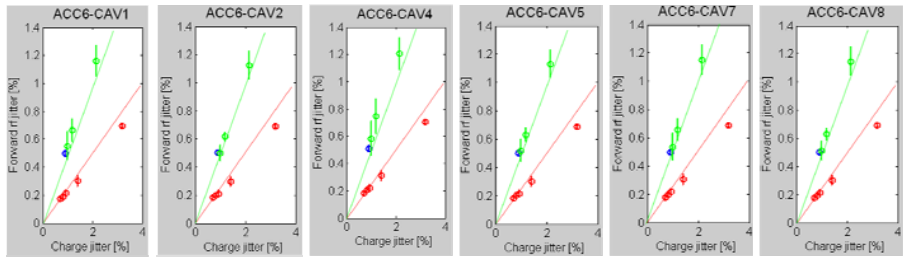

Figure 12: $2^{nd}$ input rf flat top jitter as a function of bunch charge jitter and the expected slopes (1/2 for the green solid line, 1/4 for the red). Red: 1MHz/3nC beam with piezos off; Blue: 3MHz/3nC beam with piezos off; Green: 3MHz/3nC beam with piezos on.

### 3.2  Cavity probe and reflected rf signals

Figure 13 shows the probe signal flat top jitter as a function of gradient for close-loop, beam-on operation. Note for some cavities the flat top gradient was not very flat so there is a wide spread of gradients (the cavity $Q_{ext}$ values need to be better optimized); see Figure 14 for the averaged cavity probe signals' profile in ACC6. Nonetheless, the jitter values are roughly consistent with those in the beam off case if one assumes the cavity detuning was close to optimal in most cases. Figure 15 shows ACC6 reflected rf waveforms where the reduction in the reflected power before beam turn-on (at ~ 700 µs) with piezo compensation can clearly be seen.

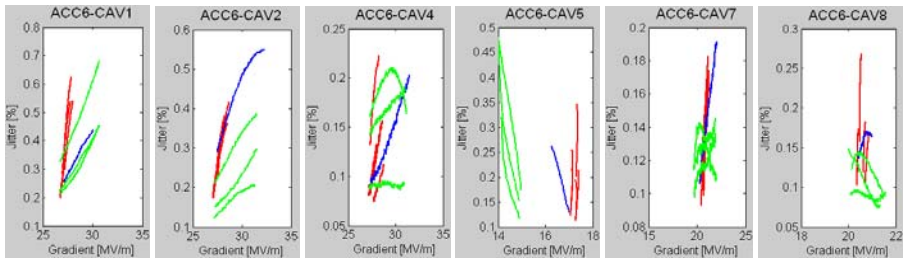

Figure 13: Probe flat top jitter as a function of gradient. Red: 1MHz/3nC beam with piezos off; Blue: 3MHz/3nC beam with piezos off; Green: 3MHz/3nC beam with piezos on.

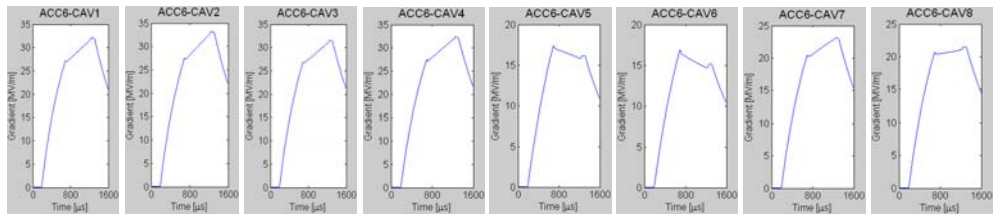

Figure 14: Cavity probe signals' profile in ACC6

LCWS/ILC 2010

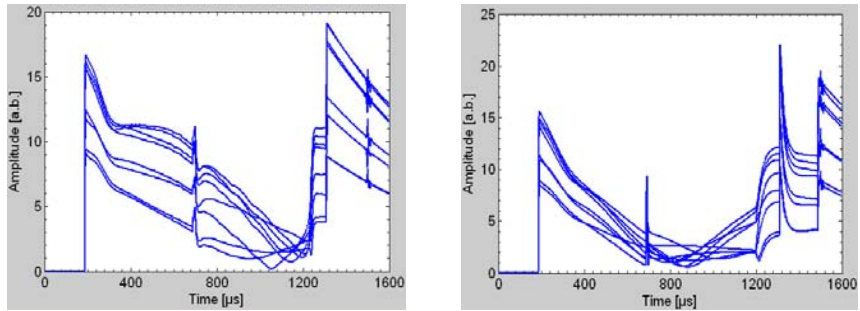

Figure 15: Reflected rf waveforms for cavities in ACC6. Left: 3MHz/3nC beam with 1600 bunches and piezos off; Right: 3MHz/3nC beam with 1500 bunches and piezos on.

## 4   Conclusions

During beam operation, the Feedback system does well to suppress cavity gradient jitter from beam loading variations. The remaining gradient jitter appears to be dominated by the effect of microphonics and a simple analytical model matches the jitter dependence on cavity detuning and gradient. Finally, piezo actuators show promise for significantly reducing the rf overhead required to compensate Lorentz force detuning.